# Fast, Approximate Synthesis of Fractional Gaussian Noise for Generating Self-Similar Network Traffic

Vern Paxson
Network Research Group
Lawrence Berkeley National Laboratory[*]
1 Cyclotron Road, Berkeley CA 94720
vern@ee.lbl.gov

## Abstract

Recent network traffic studies argue that network arrival processes are much more faithfully modeled using statistically *self-similar* processes instead of traditional Poisson processes [LTWW94, PF95]. One difficulty in dealing with self-similar models is how to efficiently synthesize traces (sample paths) corresponding to self-similar traffic. We present a fast Fourier transform method for synthesizing approximate self-similar sample paths for one type of self-similar process, Fractional Gaussian Noise, and assess its performance and validity. We find that the method is as fast or faster than existing methods and appears to generate close approximations to true self-similar sample paths. We also discuss issues in using such synthesized sample paths for simulating network traffic, and how an approximation used by our method can dramatically speed up evaluation of Whittle's estimator for $H$, the Hurst parameter giving the strength of long-range dependence present in a self-similar time series.

## 1 Introduction

When modeling network traffic, packet arrivals are often assumed to be Poisson processes because such processes have attractive theoretical properties [FM94]. Recent work, however, argues convincingly that local-area network traffic is much better modeled using statistically *self-similar* processes [LTWW94], which have much different theoretical properties than Poisson processes. A subsequent investigation suggests that the same holds for wide-area network traffic [PF95].

The strength of self-similar models is that they are able to incorporate *long-range dependence*, which informally means significant correlations across arbitrarily large time scales. For many networking questions, the presence or absence of long-range dependence plays a crucial role in the behavior predicted by analytic models. For example, the presence of long-range dependence can completely alter the tail of queue waiting times [ENW96].

The theory of self-similar stochastic processes is not nearly as well-developed as that for Poisson processes. But given the strong empirical evidence that self-similar models are much better than Poisson models at capturing crucial network traffic characteristics such as burstiness, it has become important to develop tools for understanding self-similar processes, and for generating synthetic network traffic that reflects the salient characteristics of these processes.

In this paper we present a fast algorithm for generating approximate sample paths for a type of self-similar process known as *fractional Gaussian noise* (FGN) [B92b]. The algorithm is based on synthesizing sample paths that have the same power spectrum as FGN. These sample paths can then be used in simulations as traces of self-similar network traffic. The key to the algorithm is a fast approximation of the power spectrum of an FGN process; this approximation also has application for fast estimation of the strength of long-range dependence (Hurst parameter) present in network arrival processes.

The next section defines self-similar processes and presents some of their properties and existing methods of synthesizing self-similar sample paths. These methods have drawbacks of either being computationally expensive, or generating approximate self-similar sample paths that suffer from bias in the Hurst parameter, in that the achieved Hurst parameter differs from the target Hurst parameter. The following section discusses Whittle's estimator, which is used to estimate a sample's Hurst parameter, giving the strength of the long-range dependence in the sample. In § 4 we present our Fourier transform method for synthesizing approximate FGN, and in § 5 evaluate the method in several ways to assess how well it approximates FGN. We then in § 6 discuss some issues in using synthesized FGN for simulating network traffic. § 7 presents a method for speeding up Whittle's estimator dramatically at little cost to accuracy, and § 8 summarizes our findings. In an appendix we give a program written in the *S* language for implementing our method.

---

[*]This paper appears in *Computer Communication Review* 27(5), pp. 5-18, Oct. 1997. This work was supported by the Director, Office of Energy Research, Office of Computational and Technology Research, Mathematical, Information, and Computational Sciences Division of the United States Department of Energy under Contract No. DE-AC03-76SF00098.

## 2  Self-similar processes

We begin with two definitions. A stationary process is *long-range dependent* (LRD) if its autocorrelation function $r(k)$ is nonsummable (i.e., $\sum_k r(k) = \infty$) [C84]. Thus, the definition of long-range dependence applies only to infinite time series.

The simplest models with long-range dependence are *self-similar* processes, which are characterized by hyperbolically-decaying autocorrelation functions. Self-similar and asymptotically self-similar processes are particularly attractive models because the long-range dependence can be characterized by a single parameter, the Hurst parameter $H$, which can be estimated using Whittle's procedure (see § 3 below).

More specifically, the process $\{X_t\}_{t=0,1,2,\ldots}$ is *asymptotically self-similar* if

$$r(k) \sim k^{-(2-2H)} L(k) \text{ as } k \to \infty, \qquad (1)$$

for Hurst parameter $H$ satisfying $1/2 < H < 1$ and $L$ a slowly-varying function;[1] and the process is *exactly self-similar* if [BSTW95] [C84, p.59]:

$$r(k) = 1/2 \left[(k+1)^{2H} - 2k^{2H} + (k-1)^{2H}\right].$$

For any process $\{X_t\}_{t=0,1,2,\ldots}$ we can consider an "aggregated" version $\{X_t^{(m)}\}$ constructed by partitioning $\{X_t\}$ into non-overlapping blocks of $m$ sequential elements and constructing a single element of $X_t^{(m)}$ from the average of the $m$ elements:

$$X_t^{(m)} = \frac{1}{m} \sum_{i=tm-m+1}^{tm} X_i. \qquad (2)$$

Thus $\{X_t^{(m)}\}$ corresponds to viewing the process $\{X_t\}$ using a time scale that is a factor of $m$ coarser than that used to view $\{X_t\}$ itself.

For typical stochastic processes, as $m$ increases the autocorrelation of $\{X_t^{(m)}\}$ *decreases* until in the limit the elements of $\{X_t^{(m)}\}$ are uncorrelated. For a self-similar process, on the other hand, the process $\{X_t\}$ and the aggregated process $\{X_t^{(m)}\}$ have the *same* autocorrelation function.

From these definitions it is not obvious at first glance that self-similar processes actually exist, but in fact a number of families of self-similar processes are known [ST94].

The most widely-studied self-similar processes are *fractional Gaussian noise* (FGN) and *fractional ARIMA processes* [B92b, ST94, GW94]. Associated with FGN is fractional Brownian motion (FBM), which is simply the integrated version of FGN (that is, an FBM process is simply the sum of FGN increments). In this paper we are concerned with synthesizing FGN. There are several existing methods for synthesizing sample paths for self-similar processes—see [WTLW95] for a more complete discussion and citations—but they have drawbacks:

- Consider an alternating renewal process $R(t)$ in which the on and off periods have durations from a "heavy-tailed" (e.g., Pareto) distribution. Let $S_n$ be the process constructed by multiplexing $n$ independent instances of the $R(t)$ process, where $S_n(t)$ is the number of $R(t)$ processes that are in "on" periods at time $t$. Then $S_n$ is asymptotically (as $n$ approaches $\infty$) a self-similar process [LTWW94].

  This method is particularly attractive because it matches empirical evidence of the behavior of Ethernet traffic sources [WTSW97].

  The principle difficulty with using a simulation of $S_n$ for synthesizing a self-similar process is that one must trade off speed of computation (low $n$) against the degree of agreement with a true self-similar process (asymptotically high $n$).

- Consider an M/G/$\infty$ queue model, where customers arrive according to a Poisson process and have service times drawn from a heavy-tailed distribution with infinite variance [C84, LTWW94, PF95]. In this model, $X_t$ is the number of customers in the system at time $t$, and $\{X_t\}$ is asymptotically self-similar in the sense of Eqn. 1.

  The drawback of a method based on this observation is that the process is only asymptotically self-similar, so again one must trade off length of computation for degree of self-similarity.

- A third method of synthesizing a self-similar process is the "Random Midpoint Displacement" (RMD) method [LEWW95], which works by progressively subdividing an interval over which to generate the sample path. At each division, a Gaussian displacement is used to determine the value of the sample path at the midpoint of the subinterval. Self-similarity comes about by appropriate scaling of the variance of this displacement.

  This method has the attractive property that it is fast (see below) and that it can be used to interpolate a self-similar sample path between observations made on a larger time scale. The drawbacks of the method are that it only generates an approximately self-similar process. In particular, the Hurst parameter for the sample paths tends to be larger than the target value for $0.5 < H < 0.75$, and smaller than the target $H$ for $0.75 < H < 1$, where the "target" $H$ is the value that should result if the approximations used by the method were actually exact. In addition, for a target $H = 0.5$, the sample path should correspond to white noise, but the authors found that instead it appears correlated, since the estimated $\widehat{H}$ for their synthesized

---

[1] For a slowly-varying function $L$, $\lim_{t \to \infty} L(tx)/L(x) = 1$ for all $x > 0$. Constants and logarithms are examples of slowly-varying functions.

sample paths was almost two standard deviations above $H = 0.5$.

- A fourth method involves computing wavelet coefficients corresponding to a wavelet transform of FBM. The coefficients are then used with an inverse wavelet transformation to yield sample paths of FBM [F92]. The method is only approximate because the wavelet coefficients are not independent, but it is difficult to capture their interdependence. The author of [F92] points out that the RMD method is essentially equivalent to the wavelet method for a particular (non-orthonormal) basis. Unfortunately, the paper does not include an analysis of the quality of the synthesized FBM nor the running time of an implementation of the method. A later study claims a high degree of accuracy when using the method, but the authors evaluated sample paths of only 800 points, and used a heuristic for assessing the quality of the generated FBM [SLN94].

- A fifth method, due to Hoskings, is discussed by Garrett and Willinger in [GW94]. This algorithm generates sample paths from a fractional ARIMA process, which are asymptotically self-similar. Hoskings algorithm has the very attractive property of being *exact*, but its running time is $O(n^2)$ for generating $n$ points, quite slow; the authors report that generating 171,000 points required 10 CPU hours.

- Finally, the authors of [RP94] discuss generating FBM using an approximation to the definition of FBM. Instead of computing for each new sample point the correlational contribution of all the previous sample points (which is one way of defining FBM, and results in an $O(n^2)$ algorithm for computing $n$ points), they "block" together sample points in the distant path and only compute their aggregate contribution. By using a logarithmic blocking scheme, their approximation can produce $n$ points in $O(n)$ time. While they claim their algorithm reproduces FBM accurately, their assessment of the algorithm's accuracy is terse, making it difficult to evaluate the method's promise.

Most of the synthesis methods take large amounts of CPU time. For example, [WTLW95] discusses an AR(1) method which requires 3-5 minutes to synthesize a trace of 100,000 points when running on a massively parallel computer with 16,384 processors. The first two methods mentioned above have running times on the order of CPU hours for traces of comparable length. However, it is possible that these implementations could be sped up considerably by hand optimization, so we use these figures only as "ballpark" estimates to qualitatively assess running time.

The RMD method, on the other hand, is quite fast, requiring a couple of minutes on a SPARCstation 20 to generate 260,000 points, making it much more attractive computationally. Our method is likewise fast, one of its main strengths.

Finally, we note that a drawback of some of the fast methods (RMD, inverse wavelet transformation, and our FFT algorithm) is that they must store either part of the time series (RMD method) or the entire time series (the other methods) in memory before producing any values. Consequently, they cannot be used to generate new values "on the fly."

## 3 Whittle's estimator

A key problem when studying samples of self-similar processes is estimating the Hurst parameter $H$. A "quick and dirty" approximate estimator, based on a maximum likelihood technique due to Whittle, is given by Beran [B92b, LTWW94, GW94][2]. We now give an overview of Whittle's estimator, because some of the properties of FGN processes upon which it is based are also used by our FGN synthesis method, and because a key approximation used by our method can also be used to rapidly evaluate Whittle's estimator (see § 7).

In brief, suppose $\{x_t\}$ is a sample of a self-similar process $X$ for which all parameters except $\mathrm{Var}(X)$ and $H$ are known. Let $f(\lambda; H)$ denote the power spectrum of $X$ when normalized to have variance 1, and $I(\lambda)$ the periodogram (i.e., power spectrum as estimated using a Fourier transform) of $\{x_t\}$. Then to estimate $H$, find $\widehat{H}$ that minimizes:

$$g(\widehat{H}) = \int_{-\pi}^{\pi} \frac{I(\lambda)}{f(\lambda; \widehat{H})} \, d\lambda. \tag{3}$$

If $\{x_t\}$ has length $n$, then the above integral is readily converted to a discrete summation over the frequencies:

$$\lambda = \frac{2\pi}{n}, \frac{4\pi}{n}, \ldots, \frac{2(n-1)\pi}{n}.$$

The form of this estimator relies on the fact that the periodogram ordinates $I(\lambda)$ are asymptotically independent and exponentially distributed with mean $f(\lambda; H)$ (we use this property below).

Along with the estimator one can compute $\sigma_H^2$, its variance [G93, B92a]:

$$\sigma_H^2 = 4\pi \left[ \int_{-\pi}^{\pi} \left( \frac{\partial \log f(\omega)}{\partial H} \right)^2 d\omega \right]^{-1}.$$

When synthesizing self-similar sample paths, we can then use Whittle's estimator along with $\sigma_H^2$ to determine whether our $\widehat{H}$ is acceptably close to the $H$ we intended.

An important point regarding Whittle's estimator bears repeating: it is *not* a *test* for whether a sample of a time series is consistent with long-range dependence (see [B92a] for such a test). Rather, it is an *estimator* of $H$, *given the assumption that the power spectrum of the underlying process does indeed correspond to* $f(\lambda; H)$.

---
[2]Alternative estimation techniques based on wavelet decompositions are discussed in [KK93] and [AV97].

# 4 The Fourier Transform method

In this section we present a method, based on the Discrete Time Fourier Transform (DTFT), for synthesizing fractional Gaussian noise. The strategy behind our method is taken from [F92], and can be summarized as follows. Suppose we know $f(\lambda; H)$, the power spectrum of the FGN process. Then we can construct a sequence of complex numbers $z_i$ corresponding to this power spectrum; $z_i$ is in a sense a frequency-domain sample path. We can then use an inverse-DTFT to obtain $x_i$, the time-domain counterpart to $z_i$. Because $x_i$ has (by construction) the power spectrum of FGN, and because autocorrelation and power spectrum form a Fourier pair, $x_i$ is guaranteed to have the autocorrelational properties of an FGN process, which for many purposes are its most salient characteristic.

The difficulty with this approach lies in accurately computing $f(\lambda; H)$, and in finding $z_i$ truly corresponding to the FGN power spectrum. In particular, there is no *a priori* reason to assume that the individual $z_i$ are independent, and capturing their interdependence may prove difficult. We address this difficulty below.

Because the DTFT and its inverse can be rapidly computed using the Fast Fourier Transform (FFT) algorithm, we refer to our method as an FFT method of synthesizing fractional Gaussian noise. We will not prove that the method results in true FGN, and, indeed, it does not, due to several approximations made when developing the method. But we will instead argue that the method *effectively* produces FGN. By this we mean that the sample paths produced by the method are indistinguishable (using current statistical tests) from true FGN, so for practical purposes such as simulations the sample paths can be used in lieu of true FGN with a high degree of confidence. We refer to this approach as the "quacks like a duck" approach, from the adage that if an object looks like a duck, walks like a duck, and quacks like a duck, we might as well call it a duck.

In line with this argument, there are four tests that a sample of purported FGN must pass:

- A *variance-time plot* should show that, if the sample is aggregated by a factor of $m$ (corresponding to Eqn. 2), then, asymptotically, the variance of the aggregated version falls off by a factor of $m^{-\beta}$, where $\beta = 2(1 - H)$ [LTWW94, GW94]. This is a heuristic test in the sense that the statistical properties of these plots are not known, but it is valuable because of the accompanying physical intuition: it indicates how "bursty" the sample is when viewed over progressively larger time scales. For this reason, we prefer it to other heuristic tests such as periodogram plots or R/S plots [LTWW94, GW94].

- Beran's goodness-of-fit test for long-range dependence [B92a] must indicate that the sample is consistent with long-range dependence.

- Whittle's estimator (Eqn. 3) must yield an estimated $\hat{H}$ consistent with the "true" value of $H$ used when generating the sample.

- The marginal distribution of the sample must be normal or nearly normal, since it corresponds to a Gaussian process. This can be tested using the Anderson-Darling $A^2$ omnibus test for the normal distribution [DS86, P94]. Without this test, we must question whether it is valid to use Whittle's estimator (previous item).

Both Beran's test and Whittle's estimator (Eqn. 3) are intricately tied to the estimated power spectrum of the process. For an FGN process, the power spectrum is [B86]:

$$f(\lambda; H) = \mathcal{A}(\lambda; H)\left[|\lambda|^{-2H-1} + \mathcal{B}(\lambda; H)\right] \quad (4)$$

for $0 < H < 1$ and $-\pi \leq \lambda \leq \pi$, where:

$$\mathcal{A}(\lambda; H) = 2\sin(\pi H)\Gamma(2H+1)(1 - \cos\lambda)$$

$$\mathcal{B}(\lambda; H) = \sum_{j=1}^{\infty}\left[(2\pi j + \lambda)^{-2H-1} + (2\pi j - \lambda)^{-2H-1}\right]$$

The main difficulty with using Eqn. 4 to compute the power spectrum is the vexing infinite summation in the expression for $\mathcal{B}(\lambda; H)$, for which no closed form is known. In Appendix A we discuss a general method for approximating such infinite sums, and in particular the approximation we will use is:

$$\begin{aligned}\mathcal{B}(\lambda; H) &\approx a_1^d + b_1^d + a_2^d + b_2^d + a_3^d + b_3^d \\ &+ \frac{a_3^{d'} + b_3^{d'} + a_4^{d'} + b_4^{d'}}{8H\pi}\end{aligned} \quad (5)$$

where:

$$\begin{aligned}d &= -2H - 1 \\ d' &= -2H \\ a_k &= 2k\pi + \lambda \\ b_k &= 2k\pi - \lambda\end{aligned} \quad (6)$$

We then define $\tilde{f}(\lambda; H)$ as the approximation of Eqn. 4 obtained by using Eqn. 5 for $\mathcal{B}(\lambda; H)$. We subsequently show that this approximation is good enough to pass the "quacks like a duck" criterion.

The inputs to our method are $H$, the desired Hurst parameter, and $n$, the desired (even) number of observations in the synthesized sample path. Our method proceeds as follows:

1. Construct a sequence of values $\{f_1, \ldots, f_{n/2}\}$, where $f_j = \tilde{f}(\frac{2\pi j}{n}; H)$, corresponding to the power spectrum of an FGN process for frequencies from $2\pi/n$ to $\pi$.

2. "Fuzz" each $\{f_i\}$ by multiplying it by an independent exponential random variable with mean 1. Call the fuzzed sequence $\{\hat{f}_i\}$.

We do this because when estimating a process's power spectrum using the periodogram of a sample, the power estimated for a given frequency is distributed asymptotically as an independent exponential random variable with mean equal to the actual power ([B92b, G93]; and see §3 above).

A question regarding the accuracy of our method in producing true self-similar sample paths is the degree to which this asymptotic result can be applied to a finite power spectrum without compromising the self-similarity property.

3. Construct $\{z_1, \ldots, z_{n/2}\}$, a sequence of complex values such that $|z_i| = \sqrt{\hat{f}_i}$ and the phase of $z_i$ is uniformly distributed between 0 and $2\pi$. The random phase technique, taken from [S92], preserves the power spectrum (and thus autocorrelation) corresponding to $\{\hat{f}_i\}$, but ensures that different sample paths generated using the method will be independent. It also makes the marginal distributions of the final result normal, a requirement for fractional Gaussian noise, and also for applying the Whittle procedure using an expression for $f(\lambda; H)$ corresponding to the FGN power spectrum.

One question here is *why* the phase randomization leads to what we show in the next section is a statistically verifiable Gaussian process. (We have also verified that the absence of phase randomization results in a non-Gaussian process.) It has been suggested to us that, since phase randomization makes the different frequency components independent, when the corresponding sines and cosines are added together during the inverse transform operation, the process fits with a version of the central limit theorem proved by Lindeberg [F66, p. 262]. This theorem states that if independent, differently-distributed random variables are added, the sum converges to a normal distribution, providing that some requirements are met. The first of these—that each distribution has zero-mean and finite variance—are readily met by the corresponding sines and cosines. The third condition (the "Lindeberg condition"), however, requires that the variances of the different distributions remain small compared to their total sum, and we have not verified that this condition holds. If it does, then the theorem explains why randomization leads to a normal marginal distribution.

4. Construct $\{z'_0, \ldots, z'_{n-1}\}$, an "expanded" version of $\{z_1, \ldots, z_{n/2}\}$:

$$z'_j = \begin{cases} 0, & \text{if } j = 0, \\ z_j, & \text{if } 0 < j \leq n/2, \text{ and} \\ \overline{z_{n-j}} & \text{if } n/2 < j < n. \end{cases}$$

where $\overline{z_{n-j}}$ denotes the complex conjugate of $z_{n-j}$. $\{z'_j\}$ retains the power spectrum used in constructing $\{z_i\}$, but because it is symmetric about $z'_{n/2}$, it now corresponds to the Fourier transform of a real-valued signal (again, see [S92]).

5. Inverse-Fourier transform $\{z'_j\}$ to obtain the approximate FGN sample path $\{x_i\}$.

Appendix B gives a program written in the *S* language for implementing the above method.

## 5 Evaluation of the method

We have implemented the method described in the previous section in the *S* language (see Appendix B). It is quite fast: generating a sample path of 32,768 points takes about 11 CPU seconds on a SPARCstation IPX, and 262,144 takes about 80 seconds. We have found that on numeric-intensive tasks the *S* interpreter runs about twice as fast on a SPARCstation 20 as on the IPX model, so comparing these timings with those for the RMD method given in § 2 indicates that our implementation of the FFT method runs more than twice as fast as the implementation of the RMD method used by [LEWW95]. It could easily be that the RMD method (or the FFT method) can be sped up further by some hand-tuning; however, the general conclusion that the FFT method runs quite quickly and is comparable in efficiency with the RMD method seems clear.

We then assessed how well samples produced by the method match what we would expect for FGN. For each of $H = 0.50, 0.55, \ldots, 0.90, 0.95$ we generated ten samples of 32,768 points each, corresponding to different random seeds. We then applied the four tests mentioned above: variance-time plot, Beran's goodness of fit test, Whittle's estimator, and Anderson-Darling for normal marginal distribution.

| $H$ | $\widehat{H}$ Range | Beran | Normal | V-T Plot |
|---|---|---|---|---|
| .50 | .499-.505 | √ | √ | √ |
| .55 | .547-.556 | √ | √ | √ |
| .60 | .591-.606 | √ | √ | √ |
| .65 | .647-.659 | √ | √ | √ |
| .70 | .693-.708 | √ | √ | √ |
| .75 | .745-.754 | √ | √ | sometimes low[†] |
| .80 | .794-.806 | √ | √ | sometimes low[†] |
| .85 | .842-.855 | √ | √ | usually low[†] |
| .90 | .895-.904 | √ | No[*] | usually low[†] |
| .95 | .943-.959 | √ | No[*] | always low[†] |

Table 1: Evaluation of Synthesized Fractional Gaussian Noise.

Table 1 summarizes the results of the tests. For each "true" value of $H$, the second column gives the range over the ten seeds of the $\widehat{H}$ estimate of $H$ produced by using the Whittle procedure. As noted in § 3, in addition to $\widehat{H}$, the Whittle procedure also produces a standard deviation $\sigma_H$ associated with the estimate. For our tests, $\sigma_H$ was always about 0.004.

We were thus able to test each $\widehat{H}$ to see whether it lay within two standard deviations of the actual value of $H$. This test failed four times for the 100 sample paths tested, well within the margin of error of the Whittle procedure.[3] Two of these instances were for $H = 0.60$ (one for $H = 0.65$, one for $H = 0.95$). Two failures, each with probability $p = 0.05$, out of ten samples for $H = 0.60$, is again within the margin of error. Thus, as far as Whittle's estimator is concerned, our simulated data is wholly consistent with FGN with the desired value of $H$.

A second consistency test is to check for any trends of $\widehat{H}$ being greater than $H$ or less than $H$ more often than should occur by chance. Of the 100 samples, 55 had $\widehat{H} < H$ and 45 had $\widehat{H} > H$. This variation lies within the margin of error for the null hypothesis that $\widehat{H}$ is equally likely to be larger or smaller than $H$ (i.e., no trend). When looking at fixed values of $H$, it takes $\widehat{H} < H$ or $\widehat{H} > H$ occurring 9 or 10 times for the trend to be significant (i.e., less than 5% chance of occurring by chance). This happened three times: for $H = 0.5$, 9 of the 10 samples had $\widehat{H} > 0.5$; for $H = 0.7$ and $H = 0.75$, 9 out of 10 samples had $\widehat{H} < H$. Thus for $H = 0.5$ our method appears biased toward values of $\widehat{H}$ that are slightly too high and for $H = 0.7$ and $H = 0.75$ they are slightly too low (though in all cases still within two standard deviations). We note that the failure for $H = 0.5$ is not as severe as in the case for the RMD method, since, for our method, in all cases the value of $\widehat{H}$ is comfortably within two standard deviations of $H$, while for the RMD method the authors report the values of $H$ are barely within two standard deviations. But it remains a deficiency.

Given that the RMD method has a bias towards $H = 0.75$, we also checked for separate trends for $0.55 \leq H \leq 0.65$ (skipping $H = 0.7$ due to the bias already noted above) and $0.80 \leq H \leq 0.95$. In the first case, 14 out of the 30 samples had $\widehat{H} > H$ and, in the second case, 20 out of the 40. Both of these are within the expected range. Thus our method does not appear to suffer from the same bias.

We then applied Beran's goodness-of-fit test, for which two of the samples paths failed at the 5% level, again within the margin of error.

For $H < 0.85$, all or all but one of the sample paths passed the $A^2$ test at the 5% level for normality of the marginal distribution. That they passed means that they exhibit a striking degree of normality, as the test is very sensitive to minor deviations from normality (particularly in the tails), especially for large datasets. For $H = 0.85$, two of the sample paths failed and eight passed, still within the margin of error. For

---

[3] Here and in the sequel, when we discuss a finding of $k$ events out of $n$ as being within the margin of error, we mean the following. Assume the $k$ events are independent and each has probability $p$, where $p$ depends on the exact form of the event (in this case, $p = 0.05$ since the event is "more than two standard deviations from the mean"). Then the probability that we would observe at least $k$ such events, *if indeed they are spurious*, is given by the binomial distribution for the given values of $n$, $k$, and $p$. If this probability is greater than 5%, then the finding of $k$ events might reasonably have occurred simply due to chance, and we declare the finding within the margin of error.

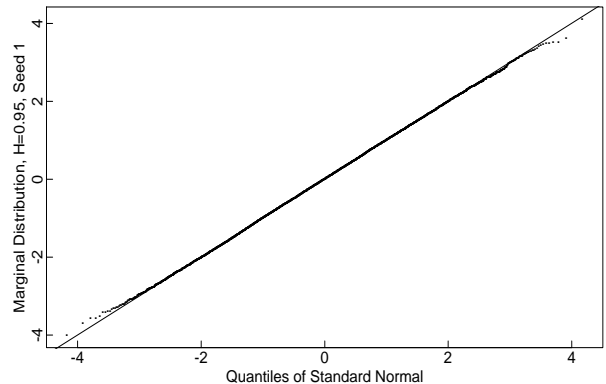

Figure 1: Q-Q Plot for Marginal Distribution of Synthesized Fractional Gaussian Noise, $H = 0.95$.

$H \geq 0.9$ the sample paths failed the $A^2$ test, but they still "look" strongly normal. For example, as shown in Figure 1, a Q-Q plot for $H = 0.95$ is indistinguishable to the eye from that of a normal distribution.

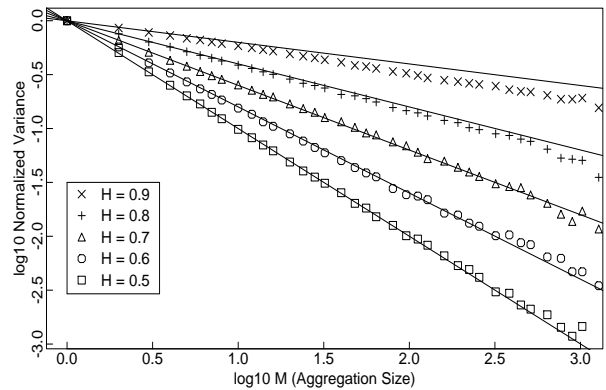

Figure 2: Variance-Time Plot for Synthesized Fractional Gaussian Noise.

The final evaluation we made was to construct variance-time plots to see whether when the sample path was aggregated by a factor of $m$ its variance fell off by the expected factor of $m^{-2(1-H)}$. Figure 2 shows such plots for one sample each of $H = 0.5, 0.6, 0.7, 0.8, 0.9$. The $x$-axis gives $\log_{10}$ of the aggregation level $m$, and the $y$-axis gives $\log_{10}$ of the (normalized) variance of the aggregated process.

The lines drawn from the origin correspond to $y = x^{-2(1-H)}$ (after a $\log_{10}$ transformation), so we expect that for a true self-similar process the variance-time plot for a given value of $H$ will coincide with the corresponding line. We found this to be the case for $H \leq 0.7$, but that for $H \geq 0.75$, the variance-time plot was sometimes, usually, or always lower (i.e., steeper-sloped) than expected, as indicated in Table 1. However, (1) the variance-time plot is based on an asymptotic relationship [C84], so this anomaly may simply be due to stronger long-range dependence (i.e., higher values of $H$) requiring longer sample paths to exhibit their true dependence; and (2) the authors of [TTW95] found that estimating $H$ from a variance-time plot introduces a bias towards underestimation of $H$, which may account for the discrepancy. Because of this latter consideration, we marked the underestimates with †'s to indicate that the problem may lie in the estimation itself, and not in the synthesis.

## 6 Application to network simulations

We have shown in the previous section that in general FGN sample paths synthesized using the FFT method pass the "quacks like a duck" criterion, in that existing statistical tools are unable to detect that the sample paths were generated using an approximate method. This finding suggests that the method can be profitably used in networking simulations.

Networking researchers wishing to simulate long-range dependent traffic face a number of issues. We comment here on some of those issues and how they relate to the FFT method.

One of the most important questions is: even if network traffic is long-range dependent, are self-similar models sufficient for capturing the long-range dependence, and if so, is the fractional Gaussian noise model an appropriate self-similar model? One way of demonstrating that self-similar models are appropriate is to show that the characteristics of network traffic match one of the theoretical models leading to self-similarity. For example, one could show that network connection characteristics match the $M/G/\infty$ queue or the heavy-tailed on/off models discussed in § 2 (see also [ENW96] and [WTSW97]). Even without compelling evidence of such a match, self-similar models still remain more attractive than traditional Poisson-based models of network arrival processes, since the latter have no long-range dependence whatsoever.

In assessing whether FGN models are appropriate, we must consider whether network arrival processes appear a close match to Gaussian processes. One important test in this regard is whether the marginal distribution of the network arrival process is close to normal, which can be assessed with a Q-Q plot (this is better than using an $A^2$ test since, as shown above, $A^2$ can be too sensitive and reject a very-close-to-normal distribution due to minor noise). We studied the wide-area link-level traces used in [PF95] to assess the degree to which the associated arrival processes have normal marginal distributions. (With our coauthor, we argued in [PF95] that these traces showed clear long-range dependence or at least "large-scale correlations.") These traces came from two sites: the Lawrence Berkeley National Laboratory's link to the external Internet (for which we examine the "PKT-4" trace below), and Digital Equipment Corporation's external Internet link, which was situated at DEC's Western Research Laboratory (the "WRL-4" trace below). The LBNL link represents a medium level of aggregation (much greater than a few sources, considerably less than a backbone), while the DEC link represents a fairly high level of aggregation.

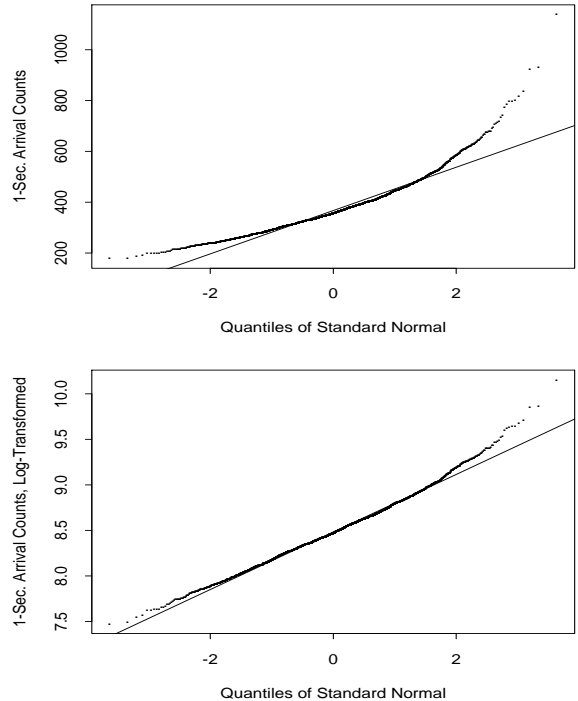

Figure 3: Q-Q Plots of Marginal Distribution for PKT-4 Arrivals (top) and Log-transformed (bottom)

For the LBNL link, we found that none of the arrival processes drawn from the traces had a marginal distribution close to normal when viewed on time scales less than 10 seconds. (Here the arrival processes are packets arrivals per fixed-duration bin.)

The authors of [LTWW94] address this problem and suggest applying a logarithmic transformation to the arrival process in an effort to pull the tails of the distribution closer to normality. We found this transformation effective, resulting in close-to-normal marginal distributions when viewing the arrival processes at time scales of 1 second, and in some cases 0.1 seconds. Figure 3 illustrates the effect of the transformation. The data in the plots is taken from [PF95]'s PKT-4 trace, which captured 1.3 million wide-area packets (one hour's worth) of all protocols. Here we have binned the trace into 1-second bins and taken the bin counts as a sample of the WAN packet arrival process. The top plot shows a Q-Q plot of the arrival process sample against quantiles of a normal distribution. The line corresponds to the plot we would expect if the sample was drawn from a normal distribution with the same mean and variance. The bottom plot shows the same sort of Q-Q plot after applying a $\log_2$-transformation to the bin counts. The fit is clearly much better, though still not exact in the tails.

The need for such a transformation suggests that, for the regime of medium levels of aggregation, either there is *not* an underlying physical process that gives a fundamental FGN characteristic to network arrivals, or that, if there is such a process, it is only a partial description and must be supplemented with additional elements (such as short-range dependence) to explain the departure from normality. This does not rule out, however, that FGN might serve as a good *approximation* for (log-transformed) network processes, or that it might still serve as a close match to more highly-aggregated traffic (next paragraph). And, of course, there might be physical processes leading directly to other self-similar models.

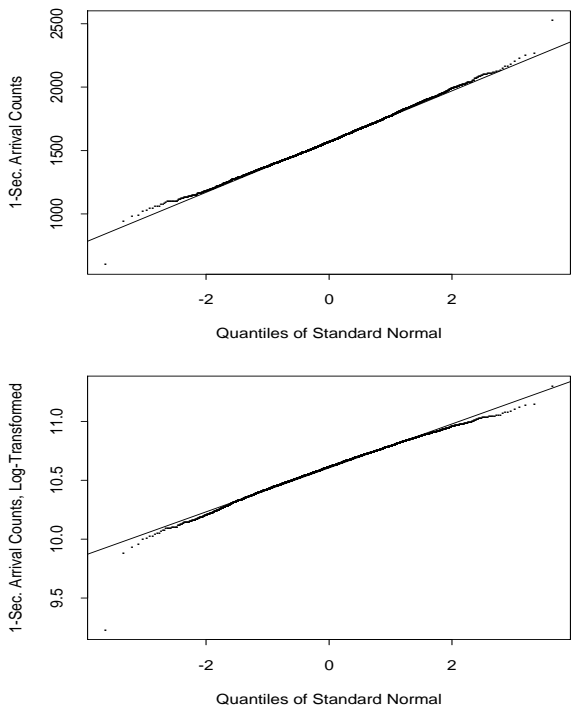

Figure 4: Q-Q Plots of Marginal Distribution for WRL-4 Arrivals (top) and Log-transformed (bottom)

On the other hand, for the more highly-aggregated DEC link, we find that the marginal distribution is much closer to normal, even without using a log-transform. Figure 4 shows the same pair of plots for the WRL-4 dataset, which captured 5.6 million wide-area packets, more than four times as many as in PKT-4. Clearly, the match to a normal marginal distribution is quite good, and is actually better than that obtained after a log-transform. This suggests that for high levels of aggregation, the normality requirement does not present any difficulty to using FGN to model packet arrivals. Furthermore, [LTWW94] found that the fit of self-similar models becomes progressively better (less asymptotic and more exact) for higher levels of traffic aggregation, suggesting that the log-transform required in Figure 3 is an indication that PKT-4 is less exactly self-similar than WRL-4.

Let us now consider the general question of transforming FGN sample paths generated by the FFT (or another) method. For FGN, the mean, variance, and Hurst parameter are all independent parameters. Thus, an FGN sample path can be scaled by a linear transformation (which preserves $H$) to achieve any desired mean and variance. In particular, the FFT method as given in Appendix B generates a zero-mean sample path, so it is replete with negative values, which are non-physical for arrival processes. If, however, one has a desired mean and variance in mind when generating the traffic, then applying the corresponding transformation should result in all or almost all positive values. If it does not, then the validity of modeling the arrival process using a Gaussian process with the given mean and variance becomes suspect.

The need in some of the situations discussed above for a logarithmic transformation, however, suggests an alternate way for converting the FGN sample path to a representation consistent with a physical process, namely the transformation $y_i = 2^{x_i}$. This transformation both preserves $H$ [G93] (so the transformed process remains long-range dependent) and results in a physical arrival process that, as shown above, in some cases more closely matches measured (wide-area) network traffic. $\{y_i\}$ does not, however, correspond to an FGN sample path anymore, and this must be kept in mind when analyzing its properties.[4]

A second issue is that often what is of interest for network simulations are interarrival times (for example, for queueing studies) and not arrival counts per bin. Since it is the arrival process that is long-range dependent and not the interarrival process, we need some way to convert arrival counts to interarrivals. First one converts the real-valued arrival process to integer counts, and then one must convert the given number of arrivals per bin into interarrivals occurring during that bin. Simple ways of distributing the arrivals over the bin are to distribute them uniformly (corresponding to exponential interarrivals) or with constant interarrivals [LEWW95]. One would expect these methods, however, to underestimate burstiness [PF95].

Another option is to use the RMD method to interpolate further self-similar sample paths within each bin. This approach makes sense if the number of arrivals is large. At some point, however, the number of arrivals is small enough that further interpolation becomes problematic, and perhaps incorrect inasmuch as the arrival process at such fine time scales may no longer be self-similar [LEWW95].

An alternative approach relates to a third issue, which is the presence of short-range dependence (SRD) in network arrival processes. In general, on small (e.g., 0.01 seconds) time scales SRD can dominate network arrivals, leading to traffic which is only asymptotically self-similar [LTWW94]. While the presence of LRD can have a dramatic effect on queueing, SRD can also significantly effect queueing behavior [ENW96]. The need to incorporate SRD into simulated network traffic suggests that one should look for ways of dis-tributing individual arrivals within a bin in such a way as to introduce SRD. For example, perhaps ARMA techniques can be applied on a bin-by-bin basis (perhaps with matching across bin boundaries) to introduce the desired level of SRD.

---

[4]The logarithmic transformation brings about another problem: since $y_i$ is always positive, it is impossible to generate an arrival count of zero. As discussed shortly, however, one must convert the real-valued sample path to an integer count anyway; incorporating rounding into this conversion will provide a mechanism for generating zero counts.

Better still would be a method of synthesizing self-similar sample paths that consistently integrates the presence of LRD and SRD. One such method, based on the Haar wavelet transform, is discussed by Kaplan and Kuo in [KK94]. At the moment their method is somewhat limited due to difficulties in parameter estimation, but still appears promising. In general, we believe wavelet methods hold great promise for characterizing and synthesizing self-similar traffic, due to the natural match between the notion of "scaling" in a wavelet transform and the notion of "invariance across different scales" in a self-similar process. Furthermore, wavelet transforms and inverse transforms can be done in $O(n)$ time, while the FFT method is limited to $O(n \log n)$, so in principle wavelet methods should also prove more efficient. See [AV97] for an extensive discussion of using wavelet techniques for analyzing LRD network traffic.

# 7 Application to fast Whittle estimation

In this section we turn to the problem of efficiently estimating $\widehat{H}$ for a given sample. While the form of Whittle's estimator given in Eqn. 3 is fairly simple, it involves evaluating $f(\lambda; H)$ (Eqn. 4), the power spectrum of the self-similar process from which the sample is presumed to have been drawn. As discussed above, for FGN exact evaluation of $f(\lambda; H)$ is an open problem, and instead one must turn to estimates. Our S program for doing Whittle estimation (written by J. Beran) addresses this problem by summing the first 200 terms of the summation expression for $\mathcal{B}(\lambda; H)$.

As shown in Appendix A, this is a good approximation, but it is slow to compute. Given that Eqn. 5 appears to be a good approximation to $\mathcal{B}(\lambda; H)$, at least for synthesizing FGN, we might then wonder whether the corresponding approximation to $f(\lambda; H)$ might be useful for computing Whittle's estimator more quickly.

To explore this possibility, we devised a modified Whittle's estimator (the estimates of which we will label $\widehat{\mathcal{H}}$) and ran it against the sample paths evaluated in the previous section. In all cases, $\widehat{\mathcal{H}}$ was within $\sigma_H$ of $\widehat{H}$, with a maximum difference between the two of 0.0028. Furthermore, we computed $\widehat{H}$ and $\widehat{\mathcal{H}}$ to a tolerance of 0.001. That is, the minimization corresponding to Eqn. 3 stopped when it found $\widehat{H}_1$ and $\widehat{H}_2$ bracketing a local minimum for which $|\widehat{H}_2 - \widehat{H}_1| \leq 0.001$. Fully 75 of the 100 samples had $|\widehat{\mathcal{H}} - \widehat{H}| < 0.001$, indicating a high degree of accuracy. In 60 of the 100 samples, $\widehat{\mathcal{H}}$ was greater than $\widehat{H}$, indicating a clear (but slight) bias towards higher values of $\widehat{\mathcal{H}}$. But for those samples with $|\widehat{\mathcal{H}} - \widehat{H}| \geq 0.001$, 14 of the 25 had $\widehat{\mathcal{H}}$ greater than $\widehat{H}$, well within the margin of error and suggesting that the slight bias might be of no practical consequence.

While the differences between $\widehat{\mathcal{H}}$ and $\widehat{H}$ are slight, the differences in running time are dramatic: using the original form of Whittle's estimator required on average about 6,500 CPU seconds on a SPARCstation IPX, while the modified estimator required about 120 CPU seconds, a savings of over a factor of 50. We conclude that using Eqn. 5 to approximate the power spectrum buys significant performance gains at only a slight cost of accuracy.

Finally, we note that we have not tested the agreement between $\widehat{\mathcal{H}}$ and $\widehat{H}$ for tolerances less than 0.001; it is possible that the agreement continues to be good, or that at finer levels limitations in the approximation of the power spectrum result in slightly inaccurate values of $\widehat{\mathcal{H}}$. Even in the latter case, one can still speed up computation of $\widehat{H}$ using our approximation as follows: perform the initial part of the minimization in Eqn. 3 using $\widehat{\mathcal{H}}$, to a tolerance of 0.005 (say); at that point switch to the more accurate but computationally expensive $\widehat{H}$ method, until achieving the desired accuracy. Thus, the approximation serves to rapidly find "the right ballpark," after which additional precision is bought with more lengthy and exact computation.

# 8 Summary and future work

One of the general problems network researchers face is how to synthesize "authentic" traffic for use in simulations and analysis. We have presented the principles behind an FFT-based method for synthesizing approximate sample paths corresponding to fractional Gaussian noise (FGN), the simplest self-similar process. We then showed that an implementation of the method is both as fast or faster than existing techniques and generates sample paths that in most regards are indistinguishable using current statistical techniques from true FGN. In particular, the FFT method appears to suffer from less bias than the Random Midpoint Displacement method, the other fast algorithm of which we are aware, though it is not completely free of bias.

Furthermore, the approximation used by the FFT method also has applications to fast evaluation of Whittle's estimator. We found that speedups by a factor of 50 were possible with only a slight loss of accuracy. Results outlined in Appendix A suggest that even this slight inaccuracy can be avoided by minor adjustments to the approximation, though we have not fully evaluated this possibility.

Section 6 raised three key issues that must be addressed when using synthesized traces for network simulations: the need to match the marginal distribution of actual traffic; the need to convert arrival counts into interarrival times; and the problem of incorporating short-range dependence into the synthesized trace. In the remainder of the section we expand on these points and related, open questions.

Suppose a group of researchers wish to perform simulation studies of network traffic, and that they accept the FFT method as an adequate mechanism for generating fractional

Gaussian noise. We believe that, in addition to the issues discussed in § 6, they need to address at least the following points:

- To what degree are they confident that the network process of interest is long-range dependent, and not simply non-stationary? This question is crucial because non-stationarity can exhibit itself in ways that look remarkably similar to long-range dependence (in particular, what appears to be strong low-frequency components, which can lead to Whittle estimates of $H > 0.5$, and variance-time plots with shallow slopes). The authors of [LTWW94] took considerable care to rule out non-stationarity effects as an explanation of long-range dependence in LAN traffic. Reference [PF95] presents an argument that time scales of 1-2 hours are stationary with regard to TCP connection arrivals, but further work is needed to convincingly rule out non-stationarity influences in WAN traffic on those time scales.

  A basic test for stationarity here is to split the dataset into two halves, and estimate $H$ independently for each half. The two estimates should, within their margins of error, yield comparable values; otherwise, it appears that $H$ is varying with time, and hence that the underlying process is non-stationary.

- Is long-range dependence a property of the traffic *sources*, or only of the traffic as seen aggregated upon the network link? The difficulty here is that different traffic sources *interact* with one another, essentially competing for a fixed resource, namely the link bandwidth. This is particularly true for TCP traffic, the dominant source of wide-area traffic today, due to TCP's adaptive window mechanism. These interactions will lengthen the "on" times during which connections transmit traffic, while also tending to homogenize the rates at which they transmit. Both of these effects will strengthen the match between network traffic and the heavy-tailed on/off model for generating self-similar traffic discussed in § 2.

  Thus one must use caution in assuming that traffic sources are well modeled using self-similar processes. Similarly, in some situations a traffic "source" might actually be traffic aggregated on a previous, upstream link. It may be tempting to model the upstream traffic as a self-similar source; but because the traffic will be further distorted by network dynamics, such a model may prove incomplete even if it is known that the traffic ultimately measured on the upstream link is indeed self-similar.

  Furthermore, even if a traffic source is self-similar, and the resulting link-level traffic is self-similar, it is possible that the relationship between the two is complex: network dynamics may significantly alter the mean, variance, Hurst parameter, and character of short-range dependence in the source. That is, the *entire* process might alter.

  Both the [LTWW94] and the [PF95] studies analyze network traffic at the link level, and thus do not provide strong guidance for how sources should be modeled. The authors of [GW94], on the other hand, show that video sources *should* be modeled as self-similar, and [WTSW97] presents compelling evidence that Ethernet sources behave in a heavy-tailed on/off fashion, which they prove yields asymptotic self-similarity when aggregated.

  No studies have yet been made on the effects of network dynamics on distorting traffic; such a study holds great promise for deepening our understanding of networks. A good starting point might be to analyze a traffic trace to characterize packet loss patterns and the resulting TCP adaptations. If, for example, TCP traffic is shaped more by the (fixed) receiver window than the (adaptive) congestion avoidance mechanisms, then it is likely that network dynamics play a minor role in contributing to self-similarity. But if, on the contrary, TCP traffic (especially large transfers, as they contribute the most to long-range dependence) is primarily shaped by congestion avoidance, then it is vital to include TCP effects when simulating networks. We show in [P97] that packet loss is not a rare event in the Internet, so we would expect that often TCP transfers are indeed operating in the congestion avoidance regime.

- Related to the previous point, if the goal of synthesize self-similar traffic is to use it as *background traffic* against which, e.g., a new transport protocol is assessed, then one must recognize that due to network dynamics it may not be possible to cleanly separate the background traffic from the introduced traffic. For example, suppose the introduced traffic attempts to aggressively use spare bandwidth as it becomes available. It may be quite unrealistic to assume that the rate of background traffic is not affected by the resulting changes in available bandwidth.

  *How* to incorporate such changes into the background traffic remains an open problem (that is, can the FFT or RMD methods be modified to extrapolate altered traffic after a change is introduced to the traffic parameters?). Indeed, as related in the previous item, even understanding *what* changes need to be incorporated, e.g., modified mean, variance, and Hurst parameter, is an open problem.

- It is crucial to understand the relative importance of an arrival process's short-range dependence versus its long-range dependence. There is no fixed balance between the two; for some situations SRD may dominate, for others LRD, and for still others each might contribute different important effects. For example, when performing a queueing simulation using a finite queue buffer, the strength of SRD in the packet arrivals might play a significant role in the delay distribution, while the strength of LRD greatly influences the packet drop patterns. The authors of [GW94] emphasize this point in a queueing simulation of video traffic by showing that the value of the Hurst parameter $H$ is necessary but not sufficient for characterizing the burstiness of the video source.

In summary, we view the FFT method not as a final answer to simulating self-similarity in network traffic, but simply as a promising starting point.

## 9 Acknowledgements

I am indebted to Sally Floyd, Steve McCanne, Murad Taqqu, Walter Willinger, and Johan Bengtsson for many productive discussions concerning this work. In particular, Sally was instrumental in shaping § 8; Steve and Johan were very patient in clarifying both fundamentals and fine points regarding Fourier and wavelet transforms; Murad helped with analyzing the underlying mathematics; and Walter encouraged the work greatly by pointing out both its promise and its pitfalls. Walter also helped with numerous technical issues, and made available Jan Beran's S programs for computing Whittle's estimator and Beran's goodness-of-fit test.

I would also like to thank Domenico Ferrari, Bill Johnston, Stefano Colpo, and several anonymous reviewers for their helpful comments.

## A  Approximating infinite sums

Since no closed form is known for the expression $\mathcal{B}(\lambda; H)$ in Eqn. 4, for our method we must instead find a suitable approximation.

Suppose $f(x)$ is a monotone decreasing function for which $\sum_{i=1}^{\infty} f_i$ converges. Then, provided the integrals exist, we have:

$$\int_1^{\infty} f(x)\,dx \leq \sum_{i=1}^{\infty} f_i \leq \int_0^{\infty} f(x)\,dx.$$

Without additional information regarding the behavior of $f(x)$, we might then use the midpoint of these integrals as an approximation for the infinite sum, since in a mean-error squared sense it is likely to be a better approximation than either the upper or the lower bound:

$$\sum_{i=1}^{\infty} f_i \approx \frac{\int_0^{\infty} f(x)\,dx + \int_1^{\infty} f(x)\,dx}{2}$$
$$\approx \frac{1}{2}\int_0^1 f(x)\,dx + \int_1^{\infty} f(x)\,dx.$$

We can further improve this approximation by explicitly retaining $k$ of the first terms of the summation:

$$\sum_{i=1}^{\infty} f_i \approx \sum_{j=1}^{k} f_j + \frac{1}{2}\int_k^{k+1} f(x)\,dx + \int_{k+1}^{\infty} f(x)\,dx. \quad (7)$$

With this addition, the approximation can be made arbitrarily close by increasing $k$.

Applying Eqn. 7 to Eqn. 4, we then have:

$$\widetilde{\mathcal{B}}_k(\lambda; H) = \sum_{j=1}^{k}(a_j^d + b_j^d) + \frac{a_k^{d'} + a_{k+1}^{d'} + b_k^{d'} + b_{k+1}^{d'}}{8\pi H}$$

where $d$, $d'$, $a_k$, $b_k$, and the like are defined as in Eqn. 6. Computationally, a great attraction of this expression is that for a given $k$ the summation can be "unrolled" and the resulting expression is then amenable to fast evaluation for a vector of different $\lambda$'s. Since the $S$ language is vector-based, this means (even on a uniprocessor) it can efficiently evaluate $\widetilde{\mathcal{B}}_k(\lambda; H)$.

The one remaining question when using this approximation is what value of $k$ to use for the best trade-off between accuracy and computational speed. We first performed the evaluations discussed in § 5 for $k = 0$ and $k = 1$, but found that the resulting $\widehat{H}$ estimates were either always lower than the target value ($k = 0$), or nearly always ($k = 1$). We skipped $k = 2$ since the assessment procedure is lengthy (requiring several CPU days for the Whittle estimations), and found $k = 3$ provided a satisfactory approximation. It is possible that $k = 2$ also performs satisfactorily, and it would run a bit faster.

One might also wonder about using the asymptotic form for the power spectrum given in [ST94], which is $h(\lambda) \sim k\lambda^{1-2H}$. This was the first form we tried, but, like $k = 0$, it resulted in $\widehat{H}$ estimates that were always too low.

We now make a brief assessment of the error introduced by using $\widetilde{\mathcal{B}}_3(\lambda; H)$ as an approximation for $\mathcal{B}(\lambda; H)$. For different values of $\lambda$ and $H$ we computed "near exact" values for $\mathcal{B}(\lambda; H)$ by summing the first 10,000 terms of the summation in Eqn. 4. We then compared these values to those obtained using $\widetilde{\mathcal{B}}_3(\lambda; H)$, and also when summing only the first 200 terms (which is what our Whittle estimation procedure uses). We refer to this latter approximation as $\mathcal{B}_{200}(\lambda; H)$.

Figure 5 shows the relative error when using the $\widetilde{\mathcal{B}}_3(\lambda; H)$ approximation. Here the relative error is:

$$\text{rel. error} = \frac{\widetilde{\mathcal{B}}_3(\lambda; H) - \mathcal{B}(\lambda; H)}{\mathcal{B}(\lambda; H)}.$$

We see that in all cases the error is less than 0.5%. We also note that $\widetilde{\mathcal{B}}_3(\lambda; H)$ is always larger than $\mathcal{B}(\lambda; H)$, suggesting that perhaps a simple adjustment can be made to the approximation to reduce much of the error; we return to this point below.

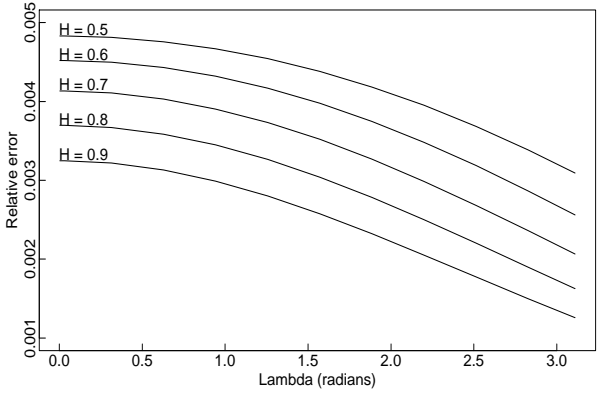

Figure 5: Relative Error in Using $\widetilde{\mathcal{B}}_3(\lambda; H)$ Approximation.

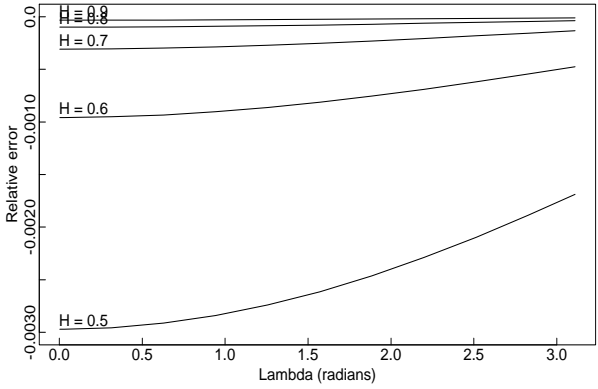

Figure 6: Relative Error in Using $\mathcal{B}_{200}(\lambda; H)$ Approximation.

Figure 6 shows a similar plot comparing summing 200 terms with summing 10,000 terms. We see that (as expected), summing only 200 terms consistently results in underestimation of $\mathcal{B}(\lambda; H)$, perforce since the summation terms are all positive. In general, using 200 terms instead of 10,000 results in little error, except for small values of $H$. Since in this case $\mathcal{B}_{200}(\lambda; H)$ *underestimates* $\mathcal{B}(\lambda; H)$ and $\widetilde{\mathcal{B}}_3(\lambda; H)$ *overestimates* $\mathcal{B}(\lambda; H)$, the error between the two approximations is particularly large for $H = 0.5$, ranging as high as 0.8%. This discrepancy may account for some of the purported bias towards slightly high values of $\widehat{H}$ for $H = 0.5$, as reported in § 5.

As noted above, the fact that $\widetilde{\mathcal{B}}_3(\lambda; H)$ is consistently larger than $\mathcal{B}(\lambda; H)$ suggests that some simple fitting might improve the approximation. We first fitted the mean absolute error $\widehat{\mathcal{B}}_3(\lambda; H) - \mathcal{B}(\lambda; H)$ as a function of $H$ (since the absolute error is relatively invariant with respect to $\lambda$), resulting in the following correction:

$$\widetilde{\mathcal{B}}_3(\lambda; H)' = \widetilde{\mathcal{B}}_3(\lambda; H) - 2^{-7.65H - 7.4}.$$

With this change, the relative error dropped from a maximum of 0.5% (as shown in Figure 5) to 0.025%, an improvement of a factor of 20. The positive bias has also disappeared, of course. We then added a correction for linear variation in $\lambda$:

$$\widetilde{\mathcal{B}}_3(\lambda; H)'' = [k_1 + k_2\lambda]\,\widetilde{\mathcal{B}}_3(\lambda; H)',$$

where:

$$\begin{aligned} k_1 &= 1.0002 \\ k_2 &= -0.000134. \end{aligned}$$

Doing so further reduces the error to 0.0075%, a factor of 3 improvement. This is within a factor of 2.5 of the relative error for $\mathcal{B}_{200}(\lambda; H)$ for $H = 0.9$ (see Figure 6), and significantly better than $\mathcal{B}_{200}(\lambda; H)$ for $H \leq 0.7$, suggesting that $\widetilde{\mathcal{B}}_3(\lambda; H)''$ could be profitably used for fast, accurate Whittle estimation. To this end, further evaluation of $\widetilde{\mathcal{B}}_3(\lambda; H)''$ remains to be done.

# B  A program for generating self-similar traces

Here is a set of S functions for implementing the method described in this paper. This program and a translation of it into C (written by Christian Schuler) are available from the Internet Traffic Archive: *http://www.acm.org/sigcomm/ITA/*

```
ss.gen.fourier <-
function(n, H)
{
# Returns a Fourier-generated sample path
# of a "self similar" process, consisting
# of n points and Hurst parameter H
# (n should be even).

  n <- n/2
  lambda <- ((1:n)*pi)/n

  # Approximate ideal power spectrum.
  f <- FGN.spectrum(lambda, H)

  # Adjust for estimating power
  # spectrum via periodogram.
  f <- f * rexp(n)

  # Construct corresponding complex
  # numbers with random phase.
  z <- complex(modulus = sqrt(f),
               argument = 2*pi*runif(n))

  # Last element should have zero phase.
  z[n] <- abs(z[n])

  # Expand z to correspond to a Fourier
  # transform of a real-valued signal.
  zprime <- c(0, z, Conj(rev(z)[-1]))
```

```
  # Inverse FFT gives sample path.
  Re(fft(zprime, inv=T))
}

FGN.spectrum <- function(lambda, H)
{
# Returns an approximation of the power
# spectrum for fractional Gaussian noise
# at the given frequencies lambda and
# the given Hurst parameter H.

  2 * sin(pi*H) * gamma(2*H+1) *
  (1-cos(lambda)) *
  (lambda^(-2*H-1) +
   FGN.B.est(lambda, H))
}

FGN.B.est <- function(lambda, H)
{
# Returns the estimate for
# B(lambda,H).

  d <- -2*H - 1
  dprime <-  -2*H
  a <- function(lambda,k) 2*k*pi+lambda
  b <- function(lambda,k) 2*k*pi-lambda
  a1 <- a(lambda,1)
  b1 <- b(lambda,1)
  a2 <- a(lambda,2)
  b2 <- b(lambda,2)
  a3 <- a(lambda,3)
  b3 <- b(lambda,3)
  a4 <- a(lambda,4)
  b4 <- b(lambda,4)
  a1^d+b1^d+a2^d+b2^d+a3^d+b3^d +
  (a3^dprime+b3^dprime +
   a4^dprime+b4^ dprime)/(8*pi*H)
}
```